\documentstyle[twoside,espcrc2}]{article}
\newcommand {\ignore}[1]{}

\newcommand{\bc}{\begin{center}}
\newcommand{\ec}{\end{center}}

\def\ifmath#1{\relax\ifmmode #1\else $#1$\fi}
\def\3quarter{{\textstyle{3 \over 4}}}

\def\ra{\rightarrow}

\def\lf{\leaders\hbox to 1em{\hss.\hss}\hfill}

\def\21{$SU(2) \ot U(1)$}
\def\ne{\hbox{$\nu_e$ }}
\def\nm{\hbox{$\nu_\mu$ }}
\def\nt{\hbox{$\nu_\tau$ }}
\def\ns{\hbox{$\nu_{sterile}$ }}
\def\nx{\hbox{$\nu_x$ }}
\def\Nt{\hbox{$N_\tau$ }}
\def\ns{\hbox{$\nu_S$ }}

        \def\etc{\hbox{\it etc. }}
        
\def\etal{\hbox{\it et al., }}

\def\rh{\hbox{right-handed }}

\def\gau{\hbox{gauge }}

\def\sm{\hbox{standard model }}

\def\neu{\hbox{neutrino }}
\def\sa{\hbox{such as }}

\def\neus{\hbox{neutrinos }}

\def\neusp{\hbox{neutrinos. }}

\def\eq#1{{eq. (\ref{#1})}}

\def\fig#1{{Fig. (\ref{#1})}}

\def\VEV#1{\left\langle #1\right\rangle}

\def\lsim{\raise0.3ex\hbox{$\;<$\kern-0.75em\raise-1.1ex\hbox{$\sim\;$}}}
\def\gsim{\raise0.3ex\hbox{$\;>$\kern-0.75em\raise-1.1ex\hbox{$\sim\;$}}}

\def\bel{\begin{letter}}
\def\eel{\end{letter}}
\def\beq{\begin{equation}}
\def\eeq{\end{equation}}
\def\bef{\begin{figure}}
\def\eef{\end{figure}}
\def\bet{\begin{table}}
\def\eet{\end{table}}
\def\bea{\begin{eqnarray}}
\def\ba{\begin{array}}
\def\ea{\end{array}}
\def\bi{\begin{itemize}}
\def\ei{\end{itemize}}
\def\ben{\begin{enumerate}}
\def\een{\end{enumerate}}
\def\ra{\rightarrow}
\def\ot{\otimes}

\def\eea{\end{eqnarray}}
\def\apj#1#2#3{          {\it Astrophys. J. }{\bf #1} (19#2) #3}

\def\ib#1#2#3{           {\it ibid. }{\bf #1} (19#2) #3}

\def\nps#1#2#3{          {\it Nucl. Phys. B (Proc. Suppl.) }
                         {\bf #1} (19#2) #3}
\def\np#1#2#3{           {\it Nucl. Phys. }{\bf #1} (19#2) #3}
\def\pl#1#2#3{           {\it Phys. Lett. }{\bf #1} (19#2) #3}
\def\pr#1#2#3{           {\it Phys. Rev. }{\bf #1} (19#2) #3}

\def\prl#1#2#3{          {\it Phys. Rev. Lett. }{\bf #1} (19#2) #3}
\def\pw#1#2#3{          {\it Particle World }{\bf #1} (19#2) #3}

\def\n.c.#1#2#3{         {\it Nuovo Cim. }{\bf #1} (19#2) #3}
\def\r.n.c.#1#2#3{       {\it Riv. del Nuovo Cim. }{\bf #1} (19#2) #3}
\def\sjnp#1#2#3{         {\it Sov. J. Nucl. Phys. }{\bf #1} (19#2) #3}

\def\zetfpr#1#2#3{         {\it Z. Eksp. Teor. Fiz. Pisma. Red. }{\bf #1}
(19#2) #3}

\def\mpl#1#2#3{          {\it Mod. Phys. Lett. }{\bf #1} (19#2) #3}

\def\ppnp#1#2#3{           {\it Prog. Part. Nucl. Phys. }{\bf #1} (19#2) #3}

\def\pc{private communication}

\relax

\emergencystretch = \hsize
\begin{document}
\pagestyle{empty}
\hyphenation{the-o-ret-ical}
\hyphenation{neu-trino}
\title{Neutrino Mass Physics }
\author{Jos\'e W. F. Valle
\address{Instituto de F\'{\i}sica Corpuscular - C.S.I.C.,
Departament de F\'{\i}sica Te\`orica, Universitat de Val\`encia\\
46100 Burjassot, Val\`encia, SPAIN  \\
E-mail VALLE at vm.ci.uv.es or 16444::VALLE
}\thanks{Work supported by DGICYT grant number PB92-0084.}
}

\begin{abstract}
Present limits on \neu masses are briefly reviewed,
along with cosmological and astrophysical hints from
dark matter, solar and atmospheric neutrino observations
that suggest neutrino masses. These would imply
many possible new phenomena \sa neutrinoless
$\beta\beta$ decay,
lepton flavour violating processes \sa
	\neu oscillations,
	$\mu \ra e \gamma$,
	$\mu \ra 3 e $,
	$\mu \ra e$ conversion in nuclei,
as well as two-body decays with the emission
of a superweakly interacting spin zero particle,
called majoron, e.g. $\mu \ra e + J$.
All of these processes may occur at levels
consistent with present or planned
experimental sensitivities.
The underlying physics may also be probed
at the high energies accessible at LEP, through related
Z decay processes.
Another possible, albeit quite indirect, manifestation
of massive neutrinos is in Higgs physics. As an example
I discuss the possibility of an invisibly decaying Higgs,
a quite generic feature of majoron models where the
lepton number is spontaneously violated close to the
electroweak scale.
\end{abstract}

\maketitle

\section{LIMITS ON NEUTRINO MASSES}

No solid theoretical principle prevents neutrinos from
having mass. In fact, most attractive extensions of the
\sm require \neus to be massive. However, theory is
not capable of predicting the scale of \neu masses any
better than it can fix the masses of the other
quarks and charged leptons, say the muon.

There are several limits on \neu masses that follow
from observation. The laboratory bounds may be
summarized as \cite{PDG92}
\beq
\label{1}
m_{\nu_e} 	\lsim 10 \: \rm{eV}, \:\:\:\:\:
m_{\nu_\mu}	\lsim 270 \: \rm{keV}, \:\:\:\:\:
m_{\nu_\tau}	\lsim 31  \: \rm{MeV}
\eeq
These limits follow purely from kinematics
and have therefore the great advantage that they
are the most model-independent of the \neu mass
limits. Note that the limit on the \nt mass may
be substantially improved at a tau factory \cite{jj}.
In addition, there are limits
on neutrino masses that follow from the nonobservation of
neutrino oscillations. I address you to ref.
\cite{granadaosc} for a detailed discussion
and compilation. As opposed to the limits in \eq{1}
\neu oscillation limits are correlated ones, involving
\neu mass differences versus mixing. Thus they rely on
the additional assumption, although quite natural in
\gau theories, that massive \neus do mix.

Apart from the above limits, there is an important
one derived from the non-observation of the
${\beta \beta}_{0\nu}$ nuclear decay process i.e.
the process by which nucleus $(A,Z-2)$ decays to
$(A,Z) + 2 \ e^-$.
This lepton number violating process would arise
via \neu exchange and, although highly favoured by phase space
over the usual $2\nu$ mode, it proceeds only if the virtual neutrino
is a Majorana particle. The decay amplitude is
proportional to
\beq
\VEV{m} = \sum_{\alpha} {K_{e \alpha}}^2 m_{\alpha}
\label{AVERAGE}
\eeq
where $\alpha$ runs over the light neutrinos.
The non-observation of ${\beta \beta}_{0\nu}$
in $^{76} \rm{Ge}$ and other nuclei leads to the limit \cite{Avignone}
\beq
\label{bb}
\VEV{m} \lsim 1 - 2 \ eV
\eeq
depending on nuclear matrix elements \cite{haxton_granada}.
Even better sensitivity is expected from the upcoming
enriched germanium experiments \cite{Avignone}.
Although rather stringent, the limit in \eq{bb}
is rather model-dependent, and does not apply
when total lepton number is an unbroken symmetry,
as is the case for Dirac \neusp Even if all \neus
are Majorana particles, $\VEV{m}$ may differ substantially
from the true neutrino masses $m_\alpha$ relevant for kinematical
studies, since in \eq{AVERAGE} the contributions of
different neutrino types may interfere destructively,
similarly to what happens in the simplest Dirac \neu case,
where the lepton number symmetry enforces that
$\VEV{m}$ automatically vanishes \cite{QDN}.

The ${\beta \beta}_{0\nu}$ decay process may
also be engendered through the exchange of scalar
bosons, raising the question of which relationship the
${\beta \beta}_{0\nu}$ decay process bears
with the \neu mass.
A simple but essentially rigorous proof
\cite{BOX} shows that, in a gauge theory,
whatever the origin of ${\beta \beta}_{0\nu}$
is, it requires \neus to be Majorana
particles, as illustrated in \fig{box}.
\bef
\vspace{4cm}
\caption{${\beta \beta}_{0\nu}$ decay and Majorana neutrinos.}
\label{box}
\eef
Indeed, any generic "black box" mechanism inducing
neutrinoless double beta decay can be closed, by W exchange,
so as to produce a diagram generating a nonzero Majorana
neutrino mass, so the relevant neutrino will,
at some level, be a Majorana particle \cite{BOX}.

Gauge theories may lead to new varieties of
neutrinoless double beta decay involving
the $emission$ of light scalars, such as
the majoron \cite{GGN}
\footnote{A related light scalar
boson $\rho$ should also be emitted.}
\beq
(A,Z-2) \rightarrow (A,Z) + 2 \ e^- + J \:.
\eeq
The emission of such weakly interacting light
scalars would only be detected through their
effect on the $\beta$ spectrum.

The simplest model leading to sizeable majoron
emission in $\beta\beta$ decays involving an
isotriplet majoron \cite{GR} leads to a new
invisible decay mode for the neutral \gau
boson with the emission of light scalars,
\beq
Z \ra \rho + J,
\label{RHOJ}
\eeq
now ruled out by LEP measurements of the
invisible Z width \cite{LEP1}.

However it has been recently shown that a sizeable
majoron-neutrino coupling leading to observable
emission rates in neutrinoless double beta decay
can be reconciled with the LEP results in models
where the majoron is an isosinglet and lepton number
is broken at a low scale \cite{ZU}. An alternative
possibility was discussed in \cite{Burgess93}.
As we have heard from Klapdor and Moe at the
moment there is only a limit on the majoron
emitting neutrinoless double beta decay leading
to a limit on the majoron-neutrino coupling
of about $10^{-4}$ \cite{klapdor_wein}.

In addition to laboratory limits, there is a cosmological
bound that follows from avoiding the overabundance of
relic neutrinos \cite{KT}
\beq
\sum_i m_{\nu_i} \lsim 50 \: \rm{eV}
\label{rho1}
\eeq
This limit is also model-dependent, as it only holds
if \neus are stable on cosmological time scales.
There are many models where neutrinos decay into
a lighter \neu plus a superweakly interacting spin
zero particle, called majoron, in such a way as to
avoid the limit in \eq{rho1} \cite{fae},
\beq
\nu_\tau \ra \nu_\mu + J \:\: .
\label{NUJ}
\eeq
Lifetime estimates in seesaw type majoron models have
been discussed in ref. \cite{V}. Here I borrow the estimate
of the model of ref. \cite{ROMA}, given by curve C in \fig{ntdecay}.
Comparing this curve with the \nt decay lifetime needed
in order to efficiently suppress the relic \nt
contribution (solid line) one sees that the theoretical
lifetimes can be shorter than required. Moreover, since these
decays are $invisible$, they are consistent with all
astrophysical observations.
If, however, the universe is to have become matter-dominated
by a redshift of 1000 at the latest (so that fluctuations have
grown by the same factor by today), the \nt lifetime has
to be shorter than indicated by the dashed line in \fig{ntdecay}
\cite{ST}. Again, lifetimes below this line are possible
\footnote{However, this lifetime limit
is less reliable than the one derived from the critical
density, as there is not yet an established theory for the
formation of structure in the universe.}.

Recently Steigman and
collaborators have argued that many values of the \nt mass
can be excluded by cosmological big-bang nucleosynthesis,
even when it decays \cite{BBNUTAU}. This, however, still
leaves open a wide region of theoretically interesting
\nt lifetime-mass values for which the searches for
the new phenomena suggested here are meaningful.
\bef
\vspace{8cm}
\caption{
Estimated \nt lifetime versus observational limits.
}
\label{ntdecay}
\eef

As a result, any effort to improve present limits on
\neu masses is definitely worthwhile. These
include experiments searching for distortions
in the energy distribution of the electrons and
muons coming from decays \sa
$\pi, K \ra e \nu$, $\pi, K \ra \mu \nu$, as
well as kinks in nuclear $\beta$ decays \cite{Deutsch}.

\section{HINTS FOR NEUTRINO MASSES}

In addition to the {\sl limits} described in the
previous section, observation also provides us with
some positive {\sl hints} for neutrino masses.
These follow from cosmological, astrophysical
and laboratory observations which I now discuss.

Recent observations of cosmic background temperature
anisotropies on large scales by the COBE  satellite,
when combined with smaller scale observations
(cluster-cluster correlations) indicate the need for
the existence of a hot {\sl dark matter} component,
contributing by about 30\% to the total mass density,
i.e. $\Omega_{HDM} \sim 0.3$ \cite{cobe}.
For this the most attractive particle candidate is a
massive neutrino, \sa as a \nt of a few eV mass.
This suggests the possibility of having observable
oscillations involving the \nt in the laboratory.
With good luck the next generation of \neu
experiments \sa CHORUS and NOMAD at CERN
and the P803 experiment proposed at Fermilab
could probe these oscillations.
The region of \nm to \nt oscillation parameters that
can be explored in the CERN experiments \cite{chorus}
is shown in \fig{cernosc}.
\bef
\vspace{8cm}
\caption{
Experimental sensitivity to \ne to \nt and \nm to \nt oscillations
}
\label{cernosc}
\eef
Despite the poorer sensitivity to \ne to \nt oscillations,
these experiments will also substantially improve the present
experimental reach in the \ne to \nt channel too.

Second, the {\sl solar \neu data} collected up to now by
the two high-energy experiments Homestake and Kamiokande,
as well as by the low-energy data on pp neutrinos from
the GALLEX and SAGE experiments still pose a persisting
puzzle \cite{Davis,granadasol}. The astrophysical explanation
of the high energy data would require not only too large
a drop in the temperature of the solar core, but would also
predict wrongly the relative degree of suppression observed
in Kamiokande and Homestake \cite{Smirnov_wein}. Indeed,
if the results of the four experiments are taken together,
they require the existence of new physics in the \neu sector
\cite{NEEDNEWPHYSICS}.
The most attractive way to account for the data
is to assume the existence of \neu conversions
involving very small \neu masses $\sim 10^{-3}$ eV
\cite{MSW}. The region of parameters
allowed by present experiments is illustrated
in \fig{msw}, taken from ref. \cite{Hata} (for similar analyses,
see ref. \cite{MSWPLOT}). Note that the fits
favour the non-adiabatic over the large mixing
solution, due mostly to the larger reduction of
the $^7 $ Be flux found in the small angle region.
\bef
\vspace{8cm}
\caption{Region of solar \neu oscillation parameters
allowed by experiment}
\label{msw}
\eef

Finally, there are hints for \neu masses from studies involving
{\sl atmospheric neutrinos}. Although the predicted
absolute fluxes of \neus produced
by cosmic-ray interactions in the atmosphere
are uncertain at the 20 \% level, their
ratios are expected to be accurate to within
5 \% \cite{atmsasso}. Present observations of
contained events by Kamiokande and IMB
indicate the existence a muon deficit
\cite{atmsasso,atm}. This is not contradicted
by preliminary Soudan 2 data.
Such deficit would suggest the existence
of \neu oscillations of the type
\nm to \nx, where \nx is \ne , \nt or \ns,
a sterile neutrino.
However, no deficit has been established
in the studies of up-going muons performed
by Baksan and IMB \cite{atmsasso}. Recent
analyses of total fluxes at IMB combined
with the studies of the stopping/passing
muon ratio can be used to severely constrain
the oscillation parameters, apparently
excluding oscillations of \nm to \nt
with maximal mixing, as expected in some
theoretical models. Similar analyses have
also been performed for the case of \nm to \ns
as well as \nm to \ne channels, where
matter effects play an important role
\cite{lipari}.

Taken at face value, the above astrophysical and cosmological
observations suggest an interesting theoretical puzzle,
if one insists in trying to account for all three observations
on solar, dark matter and atmospheric \neus within
a consistent theory. Indeed, it is difficult to reconcile
these three observations simultaneously in a world with just
the three known \neus. In order to fit all the data in a
natural way a fourth \neu species is needed and, from the
LEP data on the invisible Z width, we know that such fourth
\neu must be of the sterile type \ns.

Two basic schemes have been suggested in
which the \ns either lies at the dark matter scale
(heavy \ns \cite{DARK92}) or, alternatively, at the
MSW scale (light \ns \cite{DARK92B}).

In the first case the atmospheric \neu puzzle is
explained by \nm to \ns oscillations, and the solar
\neu deficit is explained by \ne to \nt oscillations.
However, there is a clash with the bounds from
primordial big-bang nucleosynthesis.

For this reason a second class of models was
suggested where \ns is at the MSW scale so that
the solar \neu deficit is explained by \ne to \ns
oscillations, while the deficit of \nm in the
atmospheric neutrino flux is ascribed to the
presence of \nm to \nt oscillations with
nearly maximal mixing angle
\footnote{Such maximal mixing angle \nm to \nt
oscillations seem in conflict with the most
recent combined analyses of upgoing atmospheric
\neu data.}.
In this case the limits from big-bang
nucleosynthesis can be used to single out the
nonadiabatic solution to the solar \neu problem uniquely.
Another variant of this type of model was suggested in ref.
\cite{DARK92C} in which gravitational effects are invoked
in order to produce the mass splittings $\sim 10^{-6}$ eV
required in order to generate both the MSW solar \neu
conversions as well as the oscillations of atmospheric
\neusp In this case the heavy \neu mass is about
10 KeV while the lighter \neu is also quasi-Dirac
type with mass of a few eV and therefore relevant
as dark matter. Presumably such \neu could show up
in future $\beta$ decay studies
\footnote{A similar suggestion, using vacuum oscillations
instead of MSW conversions, was suggested in \cite{MV}.}.

In short, \neu masses, besides being suggested
by theory, seem to be required to fit present
astrophysical and cosmological observations.

The first possible sign of neutrino masses
is in the propagation of neutrinos. We have
seen that the next generation of accelerator
experiments at CERN may test for the existence
of \neu oscillations involving the \nt, while
the upcoming underground experiments will
clarify whether or not solar \neu oscillations
exist and also search for neutrinoless double
beta decay with improved sensitivity. How else
could \neu masses be manifest themselves at the
laboratory?

\section{LEPTON FLAVOUR VIOLATION}

Many lepton flavour violating (LFV) decays \sa
$\mu \ra e \gamma$, which are exactly forbidden
in the standard model, can exist if \neus are
massive and, in some cases, even if not!
The possible detection of such decays, like that
of \neu oscillations, would definitely signal new
physics, closely related with the properties
of the neutrinos and the leptonic weak interaction.

There can be large rates for lepton flavour
violating decays. This is the typical situation in many models
with radiative mass generation \cite{zee,Babu88}.
For example, in the models proposed to
reconcile present hints for \neu masses
there can be large rates for lepton flavour violating
muon decays $\mu \ra e \gamma$ and $\mu \ra 3 e$,
and for the corresponding tau decays.
These decay rates may easily lie within the
present experimental sensitivities and the
situation should improve at PSI or at the
proposed tau-charm factories \cite{DARK92,DARK92B}.

Alternatively, one may assume that neutrino masses arise
at the tree level, due to the exchange of hypothetical
neutral heavy leptons (NHLS) \cite{fae},\cite{BER}-\cite{CERN}.
In the simplest models of seesaw type \cite{GRS} the NHLS are
superheavy, unless fine-tunings are allowed \cite{Buch92}.
However, in other variants \cite{SST} this is not the case.
While in the minimal case the expected rate for LFV processes
is expected to be low, due to limits on \neu masses, in other
models \cite{BER}-\cite{CERN} this suppression need not be present.
As shown in \fig{mpla}, taken from ref. \cite{3E},
present constraints on weak universality violation
allow for decay branching ratios larger than the
present experimental limits, so that these already
are already probing the masses and admixtures of the
NHLS with considerable sensitivity.
\bef
\vspace{7.4cm}
\caption{Maximum estimated branching ratios for
lepton flavour violating $\mu$ decays consistent
with lepton universality. These processes may
occur even if \neus are strictly massless.}
\label{mpla}
\eef
The dashed curve corresponds to the decay
$\mu \ra e \gamma$ while the solid one is for
$\mu \ra 3e$. For M above 10 GeV the $\mu \ra e \gamma$
branching ratio exceeds the experimental limit,
and this restricts the allowed $\mu \ra 3e$
decay branching ratio, producing the kink in the
figure. Similar estimates can be done for the
corresponding tau decays \cite{3E}. The results
one finds are summarized in table 1.
\begin{table}
\begin{center}
\caption{Estimated $\tau$ decay branching ratios
consistent with observation. $J$ denotes the majoron.
}
{}.
\begin{displaymath}
\begin{array}{|c|cr|}
\hline
\mbox{channel} & \mbox{strength} & \mbox{} \\
\hline
\tau \ra \mu + J &  \lsim 10^{-3} & \\
\tau \ra e + J &  \lsim 10^{-4} & \\
\hline
\tau \ra e \gamma ,\mu \gamma &  \lsim 10^{-6} & \\
\tau \ra e \pi^0 ,\mu \pi^0 &  \lsim 10^{-6} & \\
\tau \ra e \eta^0 ,\mu \eta^0 &  \lsim 10^{-6} - 10^{-7} & \\
\tau \ra 3e , 3 \mu , \mu \mu e, \etc &  \lsim 10^{-6} - 10^{-7} & \\
\hline
\end{array}
\end{displaymath}
\end{center}
\end{table}
Clearly the allowed rare decay branching ratios
in this case lie within the sensitivities of the
planned tau and B factories. For details see
ref. \cite{TTTAU}.

Another $\mu$ violating process that may be large
is the neutrinoless muon to electron conversion
in a nucleus \cite{kosmas}
\begin{equation}
\mu ^- \, + \, (A,Z) \ra e^-\, +\, (A,Z)^*
\end{equation}
This offers a good place for studying the muon number
violation, complementary to the corresponding elementary
particle decays considered above. In this case, too,
large rates are possible without conflicting any
information from particle physics \cite{KOSMAS}.

In addition to these rare decays, there can be
others, more closely related to the masses of the
neutrinos. In any model where \neu masses arise from
the spontaneous violation of an ungauged lepton number
symmetry there is, as a result, a physical Goldstone
boson, called majoron. In order for its existence to be
consistent with the measurements of the invisible $Z$
decay width at LEP, the majoron should be a singlet under
the \21 \gau symmetry. Although the original majoron
proposal was made in the framework of the minimal seesaw
model, and required the introduction of a relatively
high energy scale associated to the mass of the \rh
\neus \cite{CMP}, there is no need for this at all.
There is a vast class of models where the lepton number
is broken close to the weak scale and which can produce
a new class of lepton flavour violating decays.
These include single majoron emission in $\mu$
and $\tau$ decays, which would be "seen" as
bumps in the final lepton energy spectrum,
at half of the parent lepton mass in its rest frame.

I will now briefly mention, as an example of this
situation, the models where supersymmetry is realized
in a \21 context in such a way that R parity is broken
spontaneously close to the weak scale \cite{MASI_pot3}.
This model (called RPSUSY, for short) contains a majoron,
as the spontaneous
violation of R parity requires the spontaneous violation
of the continuous lepton number symmetry.
Its existence leads to single majoron emitting $\mu$
and $\tau$ decays, whose allowed rates are illustrated
in \fig{muej}, taken from ref. \cite{NPBTAU}.
One sees that these majoron emitting $\mu$ decay
modes may occur at a level which lies close to
the present limit \cite{SINGLE}.
An important role is played in this estimate by
constraints related to flavour and/or total lepton
number violating processes \sa those arising from
negative neutrino oscillation and neutrinoless double
$\beta$ decay searches, as well as from the failure to
observe secondary peaks in weak decays, e.g. $\pi, K \ra \ell \nu$,
with $\ell = e$ or $\mu$.
\bef
\vspace{8cm}
\caption{Branching ratios for lepton flavour
violating $\mu$ decays with majoron emission
}
\label{muej}
\eef
The attainable branching ratios for the corresponding
tau decays $ \tau \ra \mu + J$ and $ \tau \ra e + J$
are also indicated in table 1 \cite{NPBTAU}. These
processes may also be ideally studied at a tau-charm
factory, especially in schemes with monocromator \cite{TTTAU}.

\section{SIGNALS OF NEUTRINO MASSES AT HIGH ENERGIES}

The physics of rare $Z$ decays beautifully
complements what can be learned from the
study of rare LFV muon and tau decays.
For example, if the \Nt is lighter than the $Z$
there can be new Z decays \sa
\footnote{There may also be CP violation in lepton
sector, even when the known \neus are
strictly massless \cite{CP1}. The corresponding
decay asymmetries can be of order unity with respect
to the corresponding LFV decays \cite{CP2}.} \cite{CERN},
\beq
Z \ra N_{\tau} + \nu_{\tau}
\eeq
This decay follows from the off-diagonal
neutral currents characteristic of models
with doublet and singlet lepton \cite{2227}.
Subsequent \Nt decays would then give rise to
large missing energy events, called zen-events.
As seen in table 2 this branching ratio can be
as large as $\sim 10^{-3}$ a value that is already
superseded by the good limits on such decays from
the searches for acoplanar jets and lepton pairs from $Z$
decays at LEP \cite{opal}
\begin{table}
\begin{center}
\caption{Branching ratios for rare $Z$
decays. \Nt denotes an isosinglet
neutral heavy lepton, $\chi$ denotes
the lightest chargino (electrically charged
SUSY fermion) and $\chi^0$ is the lightest neutralino.
}
{}.
\begin{displaymath}
\begin{array}{|c|cr|}
\hline
\mbox{channel} & \mbox{strength} & \mbox{} \\
\hline
Z \ra \Nt \nt &  \lsim 10^{-3} & \\
Z \ra e \tau &  \lsim 10^{-6} - 10^{-7} & \\
Z \ra \mu \tau &  \lsim 10^{-7} & \\
\hline
Z \ra \chi \tau &  \lsim 6 \times 10^{-5} & \\
Z \ra \chi^0 \nt &  \lsim 10^{-4} & \\
\hline
\end{array}
\end{displaymath}
\end{center}
\end{table}

If $M_{\Nt}>M_Z$ then \Nt can not be directly produced at
LEP1 but can still mediate rare LFV decays \sa $Z \ra e \tau$
or $Z \ra \mu \tau$ through virtual loops. Unfortunately,
the stringent limits on $\mu \ra e \gamma$ exclude any
possible detectability at LEP of the corresponding
$Z \ra e \mu$ decay. In fact, under realistic luminosity
and experimental resolution assumptions, it is unlikely
that one will be able to see even the $e\tau$ or $\mu\tau$
decays at LEP \cite{ETAU}. In any case, there have been
dedicated searches which have set good limits \cite{opalfv}.

There have been other studies of the effects of
isosinglet NHLS both at $e^+ e^-$ collisions \cite{Djouadi}
as well as hadron supercolliders \cite{Bhattacharya}.

Another rare $Z$ decay is the single production of the
lightest chargino, characteristic of the RPSUSY model
\cite{ROMA,RPCHI},
\beq
Z \ra \chi \tau
\eeq
where the lightest chargino mass is assumed to smaller than
the Z mass. As shown in \fig{5}, the allowed branching ratio
lies close to the present LEP sensitivities. The search for
this decay mode is now underway by the L3 collaboration
\cite{Felcini93}.
\bef
\vspace{7cm}
\caption{Branching ratios for single
chargino production in $Z$ decays may exceed $10^{-5}$
}
\label{5}
\eef
Similarly, the lightest neutralino could also be
singly-produced as $Z \ra \chi^0 \nu_\tau$ \cite{ROMA,RPCHI}.
Being unstable due to R parity violation, $\chi^0$ is
not necessarily an origin of events with missing energy,
since some of its decays are into charged particles.
Thus the decay $Z \ra \chi^0 \nu_\tau$ would give rise to
zen events, similar to those of the minimal supersymmetric
standard model (MSSM), but where the missing energy is
carried by the \nt. Another possibility for zen events
in RPSUSY is the usual pair neutralino production process,
where one $\chi^0$ decays visibly and the other invisibly.
The corresponding zen-event rates can be larger than
in the MSSM.
Their origin is also quite different, being tied with
the nonzero value of the \nt mass.

Moreover,
although the \nt can be quite massive, the \ne and \nm
have a tiny mass difference, that can be chosen to lie
in the range where resonant \ne to \nm conversions provides
an explanation of solar \neu data. Due to this peculiar
hierarchical pattern, one can go even further, and regard the
rare R parity violating processes as a tool to
probe the physics underlying the solar \neu deficit
in this model \cite{RPMSW}. Indeed, the rates for
such rare decays can be used in order to
discriminate between large and small mixing
angle MSW solutions to the solar \neu problem.
Typically, in the nonadiabatic region of
small mixing one can have larger rare decay branching
ratios, as seen in Fig. 5 of ref. \cite{RPMSW}.

Finally, it is possible to find manifestations of massive
\neus even at the superhigh energies available at
hadron supercolliders LHC/SSC. An example has
been discussed in the RPSUSY model, i.e. the
possible single production of supersymmetric
fermions \cite{RPLHC}.

The above examples illustrate how the search for
rare decays can be a more sensitive probe of \neu
properties than the more direct searches for \neu
masses, and therefore complementary.

\section{INVISIBLE HIGGS DECAYS}

In many models \cite{JoshipuraValle92} \neu masses
are induced from the spontaneous violation of a global
$U(1)$ lepton number symmetry by an \21 singlet vacuum
expectation value $\VEV{\sigma}$, in such a way that
$m_\nu \to 0$ as $\VEV{\sigma} \ra 0$.
In contrast with the more usual seesaw majoron model
\cite{CMP}, a low scale for the lepton number violation,
close to the electroweak scale, is {\sl preferred} in
these models, since it is required in order to obtain
small neutrino masses \cite{JoshipuraValle92}
\footnote{Another example is provided by the
RPSUSY models \cite{HJJ}.}.
Another cosmological motivation for low-scale
majoron models has been given in ref. \cite{Goran92}.

In these models, although the majoron has very tiny couplings to
matter, it can have significant couplings to the
Higgs bosons.
This implies that the Higgs boson may decay
with a substantial branching ratio into the
invisible mode \cite{JoshipuraValle92,Joshi92}
\begin{equation}
h \rightarrow J\;+\;J
\label{JJ}
\end{equation}
where $J$ denotes the majoron. The presence of
this invisible Higgs decay channel can affect
the corresponding Higgs mass bounds in an
important way, as well as lead to novel
search strategies at higher energies.

The production and subsequent decay of any Higgs boson
which may decay visibly or invisibly involves three independent
parameters: the Higgs boson mass $M_H$, its coupling
strength to the Z, normalized by that of the \sm, call
this factor $\epsilon^2$, and the invisible Higgs boson
decay branching ratio.

The results published by the LEP experiments on the
searches for various exotic channels can be used
in order to determine the regions in parameter space
that are ruled out already. The procedure was described
in \cite{alfonso}. Basically it combines the results
of the standard model Higgs boson searches with those one
can obtain for the invisible decay.
For each value of the Higgs mass, the lower bound on
$\epsilon^2$ can be calculated as a function of the
branching ratio $BR(H \rightarrow $ visible), both this
way as well as through the \sm Higgs search analyses
techniques. The weakest of such bounds for
$BR(H \rightarrow $ visible) in the range
between 0 and 1, provides the absolute bound on $\epsilon^2$.
This procedure can be repeated for each value of $M_H$, thus
providing an an exclusion contour in the plane $\epsilon^2$
vs. $M_H$, shown in \fig{alfonso2}, taken from ref. \cite{alfonso}.
The region in $\epsilon^2$ vs. $M_H$ that is already excluded by the
present LEP analyses holds {\sl irrespective of the mode of Higgs decay},
visible or invisible.
\bef
\vspace{7cm}
\caption{Region in the $\epsilon^2$ vs. $m_H$ that can be
excluded by the present LEP1 analyses, independent of the
mode of Higgs decay, visible or invisible (solid curve).
Also shown are the LEP2 extrapolations (dashed).}
\label{alfonso2}
\eef
Finally, one can also determine the additional
range of parameters that can be covered by LEP2
for a total integrated luminosity of 500 pb$^{-1}$
and centre-of-mass energies of 175 GeV and 190 GeV.
This is shown as the dashed and dotted curves in
\fig{alfonso2}.

The possibility of invisible Higgs decay
is also very interesting from the point of
view of a linear $e^+ e^-$ collider at higher
energy \cite{EE500}. Heavier, intermediate-mass,
invisibly decaying Higgs bosons can also be searched at high energy
hadron supercolliders such as LHC/SSC \cite{granada}.
The limits from LEP discussed above should
serve as useful guidance for such future searches.

\section{CONCLUSION}

Present cosmological and astrophysical observations,
as well as theory, suggest that neutrinos may be massive.
Existing data do not preclude neutrinos from being responsible
for a wide variety of measurable implications at the laboratory.
These new phenomena would cover an impressive region of energy,
from weak decays, \sa $\beta$ and double $\beta$ decays,
to neutrino oscillations, to rare processes with lepton flavour
violation, from nuclear $\mu e$ conversions and rare muon
decays, to tau decays, to Z decays. Moreover, neutrino masses
might even affect the electroweak symmetry breaking sector in
an indirect, but important way.

The next generation of experiments \sa those
looking for spectral distortions in weak decays,
\neu oscillation searches sensitive to \nt
as dark matter (CHORUS/NOMAD/P803),
$e^+ e^-$ collisions from ARGUS/CLEO, to
TCF/BMF (tau-charm and B factories), to
LEP; and finally even the next decade hadron
supercolliders LHC/SSC could all be sensitive
to \neu properties!

It is therefore quite worthwhile to keep
pushing the underground experiments, for any
possible confirmation of \neu masses.
These includes experiments with enriched germanium
looking for neutrinoless $\beta \beta$ decays,
solar \neu experiments GALLEX and SAGE,
as well as Superkamiokande, Borexino, and
Sudbury. The same can be said of the ongoing
studies with atmospheric \neusp

Similarly, a new generation of experiments capable
of more accurately measuring the cosmological
temperature anisotropies at smaller angular scales than
COBE, would be good probes of different models of
structure formation, and presumably shed further
light on the need for hot \neu dark matter.

\bibliographystyle{ansrt}

\end{document}